# Nanotribology of biopolymer brushes in aqueous solution using dissipative particle dynamics simulations: an application to PEG covered liposomes in theta solvent

A. Gama Goicochea[†,1], E. Mayoral[1], J. Klapp[1,2], and C. Pastorino[3]

[1]Instituto Nacional de Investigaciones Nucleares, Carretera México – Toluca s/n, La Marquesa Ocoyoacac, Estado de México 52750, Mexico

[2]Departamento de Matemáticas, CINVESTAV del IPN, México D. F. 07360, Mexico

[3]Departamento de Física, Centro Atómico Constituyentes, CNEA – CONICET, Av. General Paz 1499, Provincia de Buenos Aires 1650, Argentina

## Abstract

We undertake the investigation of sheared polymer chains grafted on flat surfaces to model liposomes covered with polyethylene glycol brushes as a case study for the mechanisms of efficient drug delivery in biologically relevant situations, for example, as carriers for topical treatments of illnesses in the human vasculature. For these applications, specific rheological properties are required, such as low viscosity at high shear rate to improve the transport of the liposomes. Therefore, extensive non- equilibrium, coarse – grained dissipative particle dynamics simulations of polymer brushes of various lengths and shear rates are performed to obtain the average viscosity and the friction coefficient of the system as functions of the shear rate and polymerization degree under theta – solvent conditions, and find that the brushes experience considerable shear thinning at large shear rates. The viscosity ($\eta$) and the friction coefficient ($\mu$) are shown to obey the scaling laws $\eta \sim \dot{\gamma}^{-0.31}$, and $\mu \sim \dot{\gamma}^{0.69}$ at high shear rate ($\dot{\gamma}$) in theta solvent, irrespective of the brushes degree of polymerization. These results confirm recent scaling predictions and reproduce very well trends in measurements of the viscosity at high ($\dot{\gamma}$) of red blood cells in a liposome containing medium.

---

[†] Corresponding author. Electronic mail: agama@alumni.stanford.edu

# I. INTRODUCTION

Polymer brushes, formed when polymer molecules grafted to a surface are stretched, are important because they play a key role in the stability of colloidal dispersions [1, 2], in the process of enhanced oil recovery [3], in the design and applications of stimuli – responsive materials [4], and even in the characterization of the mechanical properties of cancerous human cells [5], among other reasons [6]. They display some general fundamental properties, and scaling relations have been derived for them under circumstances such as high grafting density or negligible chain – chain interaction [7-9]. When polymer brushes are sheared a plethora of phenomena that are intrinsic to the non – equilibrium nature of the shearing appear. For example, experiments on fluids compressed between plates under the influence of external shear have shown that the viscosity of the fluid can be substantially reduced if a polymer brush is grafted to each plate [10]. Biological examples of sheared polymer brushes occur in articular cartilage surfaces [11] and synovial joints [12], in glycocalyx filaments that coat the human circulatory system [13], and in glycosylated cell surfaces and liposomes, which can be used as carriers for drug delivery [14]. Despite all these studies, the experimental understanding of the molecular mechanisms that take place in the various environments cited before is still incomplete because, among other reasons, characterizing variables such as the thickness of the polymer brushes on the sheared surfaces is still difficult to accomplish [6]. These difficulties arise partly because the self – assembly processes typically used for the modification of surfaces depend on chemical reactions that do not necessarily yield a well – defined brush length [6]. In this regard, computer simulations have come to play an increasingly important role, and there are now various works investigating the role of the polymer brush length, the spacing between the plates, the properties of the solvent and the influence of the electrostatic interactions, the fluid´s shear thinning as the shear rate is increased, and the effect of changing the polymer grafting density on the plates, to name but a few [15-24].

However, most works have been carried out for polymers in good solvent conditions, with very few exceptions [19, 21 – 23]. It is known that some biopolymers found in an aqueous environment are in the borderline between good solvent and theta solvent conditions [21, 25]. An important example is that of polyethylene glycol (PEG) polymer brushes on liposomes, which are used as carriers under flow for efficient drug delivery [26, 27]. PEG is known to have a diminishing solubility as its molecular weight increases [28], and for molecular weights in the range of 2000 to 10000 g/mol in water it has been reported to be slightly below its theta temperature, when placed at the human body temperature [29]. Since the mechanisms of drug delivery are of paramount importance for pharmaceutical design, and their understanding is still far from complete, we have undertaken here the study of the viscosity and friction of PEG brushes on surfaces that experience an external force under theta solvent conditions, for the first time. To reach scales comparable to those of typical non-equilibrium experiments we have carried out particle – based, mesoscopic molecular dynamics simulations, using the method known as dissipative particle dynamics (DPD) [30], which has been successful in predicting correctly properties of polymer brushes both in equilibrium and non-equilibrium situations [31].

In what follows we report DPD simulations of the rheological properties of biopolymer (PEG) brushes grafted to parallel flat plates (liposomes), including the solvent explicitly, which is crucial to reproduce experiments measuring friction between polymer brushes [17, 18]. We want to model the behavior of tethered proteins and biopolymers in environments such as vessels, where the separation between plates is essentially constant but allow for variations in the degree of polymerization, at fixed grafting density and substrate separation. In particular, we would like to determine the optimal conditions under which PEG brushes on liposomes promote the flow of the latter as mechanisms for efficient drug delivery. We obtain the viscosity and the friction coefficient as functions of the shear rate, and degree of polymerization. To our knowledge, this is the first report of scaling behavior for the viscosity and the friction coefficient at high shear rate of polymer brushes in theta solvent using soft potentials. Understanding of these phenomena is useful for the interpretation of several recent experiments in biological and other mesoscopic systems of current academic and industrial interest.

This article is organized as follows. In the following section we introduce the interaction model, the simulation algorithm and the details of the systems studied. In Section III the main results are presented, accompanied by their discussion. The final section is devoted to our major conclusions.

## II. MODELS AND METHODS

We have performed DPD simulations of linear grafted polymers in the canonical ensemble (fixed density and temperature). The DPD model is by now well known; therefore we shall be brief here. Three forces make up the basic DPD structure: a conservative force ($\vec{F}_{ij}^C$), that accounts for the local pressure and is proportional to an interaction constant, $a_{ij}$; a dissipative force ($\vec{F}_{ij}^D$), which represents the viscosity arising from collisions between particles, proportional to the relative velocity of the particles and to a constant, $\gamma$; and a random force ($\vec{F}_{ij}^R$), that models the Brownian motion of the particles, with an intensity given by the constant $\sigma$, all acting between any two particles, $i$ and $j$. The spatial dependence of these forces is usually chosen as shown in equations (1-3).

$$\vec{F}_{ij}^C = a_{ij}\left(1 - {r_{ij}}/{R_c}\right)\hat{e}_{ij} \tag{1}$$

$$\vec{F}_{ij}^D = -\gamma\left(1 - {r_{ij}}/{R_c}\right)^2 [\hat{e}_{ij}\cdot\vec{v}_{ij}]\hat{e}_{ij} \tag{2}$$

$$\vec{F}_{ij}^R = \sigma\left(1 - {r_{ij}}/{R_c}\right)\hat{e}_{ij}\xi_{ij}. \tag{3}$$

In equations (1-3), $r_{ij} = r_i - r_j$ represents the relative position vector between particles $i$ and $j$, $\hat{e}_{ij}$ is the unit vector in the direction of $r_{ij}$, while $v_{ij} = v_i - v_j$ is the relative velocity between particles $i$ and $j$, with $r_i, v_i$ being the position and velocity of particle $i$, respectively. The random variable $\xi_{ij}$ is generated between 0 and 1 with a Gaussian distribution, zero mean and unit variance; $R_c$ is the cut off distance, beyond which all forces are zero. The DPD beads are all of the same size, with radius $R_c$, which is set equal to 1. The constants in the dissipation and random forces are not independent, and satisfy the relation [32]:

$$\frac{\sigma^2}{2\gamma} = k_B T \tag{4}$$

which, in effect, fixes the temperature; here, $k_B$ is Boltzmann's constant. This natural thermostat that arises from the balance between the dissipative and random forces is a defining feature of the DPD model [31]. Also, the short-range nature of the DPD forces, and their linearly decaying spatial dependence, allow the use of relatively large time steps when integrating the equation of motion. The DPD particles are representations of sections of fluid rather than physical particles, and can group several atoms or molecules, making the DPD method an attractive alternative to study systems at the mesoscopic level. Note also that the forces in equations (1)-(3) obey Newton's third law, which means momentum is conserved locally, and globally, which in turn preserves any hydrodynamic modes present in the fluid. This is a feature of fundamental importance when studying non equilibrium properties of fluids, since loss of information about these modes can lead to a different phase from that obtained when they are fully accounted for [33]. The DPD interaction model has been used successfully to predict equilibrium properties of polymer melts [34], surfactants in solution [35], and colloidal stability [36], among others. For further reading, see reference [31].

All our simulations are performed in reduced units (marked with asterisks); distances are reduced with the cutoff radius, $R_c$, which for a coarse – graining degree equal to 3 water molecules per DPD particle is equal to $R_c$=6.46 Å [31], hence $r=r^*R_c$. The time step $\delta t$ is reduced with $\delta t = (mR_C^2/k_BT)^{1/2}\delta t^*$, where $m$ is the mass of a DPD particle, while energy is reduced with $k_BT$. All other quantities can be reduced through combinations of the these relations, for example, the viscosity: $\eta = \eta^*(k_BT\delta t/R_C^3)$. Using the mass of 3 water molecules per DPD particle, at room temperature, one obtains $\delta t \approx (6.3 \times 10^{-12}\text{s})\delta t^*$ and $\eta \approx (9.64 \text{ cP})\eta^*$. All our simulations are performed for $k_BT^* = m^* = R_C^* = 1$, $\delta t^*$=0.03. In addition to the fluid monomeric particles, our system contains soft parallel surfaces as models for biological membranes, and polymer chains attached to them, forming brushes. The polymers are built as linear chains of DPD beads joined by freely rotating harmonic springs [37]. The spring constant is chosen as $\kappa = \kappa^*(k_BT/R_C^2)$, where $\kappa^*$ =100 and the spring's equilibrium position as $r_{eq} = r_{eq}^* R_C$, with $r_{eq}^* = 0.7$ in all cases [38]. In our simulations we have fixed the distance between the plates, which are placed perpendicular to the *z*-direction, at a distance of $D^*=7$. Periodic boundary conditions are applied on the *xy* – plane but not in the *z* – direction, to reinforce the confinement. The equation of motion is solved using the velocity Verlet algorithm, adapted to DPD [39]. The parameters of the dissipative and random forces intensities are, respectively, $\gamma = 4.5$ and $\sigma = 3.0$, so that $k_B T^*$ = 1 (see equation (4)). The value of the conservative force intensity (see equation (1)) was set to $a_{ij} = 78.0$ for all cases, namely for the interaction between particles of the same type (solvent – solvent, monomer – monomer) and for particles of different type (solvent – monomer). The choice $a_{ij} = 78.0$ is obtained when one uses a coarse graining degree that groups 3 water molecules in a single DPD bead [35]. Since all particle – particle interactions are equal, the polymers are in a theta – solvent, as is the case for PEG – covered liposomes in the human circulatory system [29]. It has been shown that this model for PEG on colloidal surfaces can successfully predict brush scaling laws [27], as well as adsorption isotherms on metallic oxides and disjoining pressure profiles [36]. The soft membranes on which these biopolymers are tethered are modeled as linearly decaying forces that act on the ends of the simulation box (in the *z* – direction), given by the following expression:

$$F_w(z_i) = a_w \left[1 - \frac{z_i}{z_C}\right]. \tag{5}$$

In equation (5), $F_w(z_i)$ symbolizes the force exerted by the effective wall on particle *i*, whose position component in the *z* – direction is $z_i$. The intensity of the force is given by the constant $a_w$, while $z_C$ is the cutoff distance set also equal to 1, which defines the reach of this force, i.e., $F_w(z_i) = 0$ for $z_i > z_C$. Equation (5) represents a soft membrane because the maximum repulsion between the beads and the surface is $a_w$ when $z_i = 0$; harder walls can be used for other DPD applications [36]. It is known that liposomes can be up to about 3 orders of magnitude larger than their brush thickness, and the solvent [2, 40], therefore we consider planar walls only. To fix one end of the polymers on these substrates we chose their interaction equal to $a_w$=70.0 (reduced DPD units), while the rest of the polymer beads and the solvent interact with the walls with an intensity given by $a_w$=140.0. With this choice of parameters, the polymer end experiences a less repulsive force toward the walls than the rest of the polymer (and solvent) beads, and is therefore adsorbed on them. The tethered ends of the polymers remain free to move on the $xy$ – plane, of course, as occurs also for biopolymers interacting with membranes. Although this model for membrane is soft, it remains impenetrable to polymer and solvent molecules, and we assume moreover that the drug molecules have already been incorporated into the structure.

To carry out non equilibrium simulations of these systems, we apply a constant shear velocity to the polymer ends adsorbed on the walls, of equal magnitude but opposite sign for beads on different plates. This is equivalent to moving the plates in opposite direction under the influence of a fixed external force that keeps the plates moving with constant speed, known as Couette flow [19 – 23], see Figure 1. Some of the solvent monomers are seen to penetrate the PEG brushes and carry them along in their flow to produce lubrication, while there is no interpenetration of the brushes. As pointed out before, our simulations are performed in the canonical ensemble, where particle density and temperature are kept constant; previous works have shown that completely equivalent results are obtained if one performs the simulations in the grand canonical ensemble, at fixed chemical potential, volume and temperature [17]. To keep the simulations results invariant to the particular choice of $a_{ij}$ interaction parameters, we fix the global dimensionless density to 3 [35] in all the simulations reported here; this means that when the degree of polymerization of the PEG brushes ($N$) is increased, the number of solvent molecules is reduced.

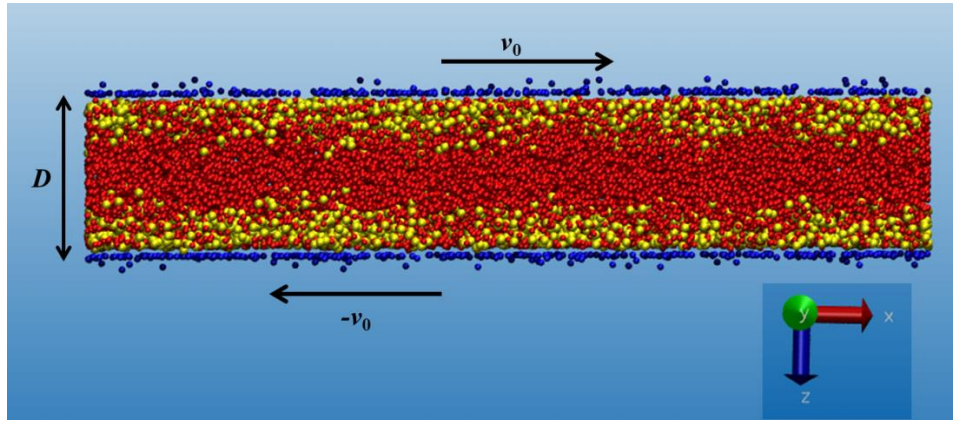

**Figure 1**. Snapshot of two linear PEG brushes made up of $N=7$ beads. The lateral dimensions of the simulation box are $L_x^*=35$, $L_y^*=35$, and the spacing between the plates is $L_z^* = D^*=7$. The grafting density is equal to $\Gamma^*=0.3$. A constant shear velocity of magnitude $v_0^*=1.0$ is applied to the tethered beads of each polymer (in dark blue); the rest of the polymer beads are shown in yellow. The solvent monomers are shown in red. Notice how the solvent penetrates the polymer brushes all the way up to the surfaces for this grafting density. See text for details.

The grafting density is defined as $\Gamma^*=N_p/A^*$, where $N_p$ is the number of polymer chains tethered on the surface and $A^*=L_x^*L_y^*$ is its reduced area. Our simulations are performed at grafting densities that lie within the brush regime ($\Gamma^*\sim 1/N$, where $N$ is the degree of polymerization, see [27]), where the "stealth" properties of the PEG brushes work best [41]. We obtain the viscosity ($\eta$), and the friction coefficient ($\mu$) of the fluid, through the relations [19]

$$\eta = \frac{\langle F_x(\dot{\gamma})\rangle/A}{\dot{\gamma}}, \quad (6)$$

$$\mu = \frac{\langle F_x(\dot{\gamma})\rangle}{\langle F_z(\dot{\gamma})\rangle}. \quad (7)$$

In equations (6) and (7) above, $\langle F_x(\dot{\gamma})\rangle$ and $\langle F_z(\dot{\gamma})\rangle$ are the mean forces that the particles on the surfaces experience along the flow direction, and perpendicularly to it, respectively; the brackets indicate an ensemble average. Equation (6) can be understood as the local definition of viscosity, applied to the entire sample. Considering a liquid confined between two planar walls, a linear flow can be generated by moving for example the top wall at constant velocity. This motion requires that a steady force be applied on the top wall, and an equal in magnitude but opposite in direction force, on the bottom wall, see Figure 1. In linear flow, the velocity gradient (shear rate, $\dot{\gamma}$) is constant throughout the liquid. The shear rate $\dot{\gamma}$ is defined as $2v_0^*/D^*$, where $v_0^*$ is the flow velocity exerted on the grafted monomers, and $D^*$ is the surface separation (see Figure 1). To obtain first a local definition of viscosity, a small cubic volume in the fluid (fluid element) can be considered. This fluid element undergoes strain, being deformed from a cubic to a parallelepiped shape, with increasing deformation in time. The amount of strain is the added lateral displacement of top and bottom planes of the fluid volume divided by its height. This strain increases at constant rate in time for linear flow, such that the strain rate i.e. the strain per unit time, is the velocity gradient in the limit of an infinitesimal fluid element. The strain rate of deformation is therefore the shear rate, $\dot{\gamma}$. On the other hand, a force acts on the top and bottom planes of the fluid element, with identical magnitude and opposite direction. The center of mass of the fluid element, suffers no acceleration and therefore the total external force on it must vanish. These forces must be proportional to the area of the top and bottom planes of the fluid element. Also, the force-per-unit-area or stress, $\sigma$, must be uniform from top to bottom in the fluid element since the fluid is not accelerating. For linear flow, this is true not only for a fluid element but for the whole fluid: every part of the fluid is under the same stress, which produces everywhere the same velocity gradient ($\dot{\gamma}$). The definition of viscosity assumes that the shear rate ($\dot{\gamma}$) produced in a fluid element is proportional to the shear stress ($\sigma$) exerted on it. The viscosity is defined as the constant of proportionality between them: $\sigma = \eta\dot{\gamma}$, or $\eta = \sigma/\dot{\gamma}$. The total shear stress on the sample ($\sigma$) can be determined from the average total force on the brush heads divided by the wall area ($F_x/A$) while the shear rate ($\dot{\gamma}$) is extracted from the slope of the linear fit of the average velocity profile. The viscosity is then expressed by equation (6). As for the friction coefficient, shown in equation (7), its calculation follows directly from the analogy with the friction coefficient between solid surfaces, namely $F_x=\mu F_z$, where $F_z$ is the force acting perpendicularly to the surface, and $F_x$ is the force acting parallel to it, which is responsible for the shear stress ($\sigma = F_x/A$) [24]. Our results were obtained from averages of simulations of up to $4\times10^3$ blocks, with each block composed of $2\times10^4$ time steps, using the first $2\times10^5$ time steps for equilibration and the rest for the production phase; when properly dimensionalized this represents a time observation window of 0.12 ms. A typical density profile of the solvent and polymer monomers is shown in Figure 2, where it is clear that the solvent penetrates the PEG brushes and reaches the surfaces, at that given grafting density.

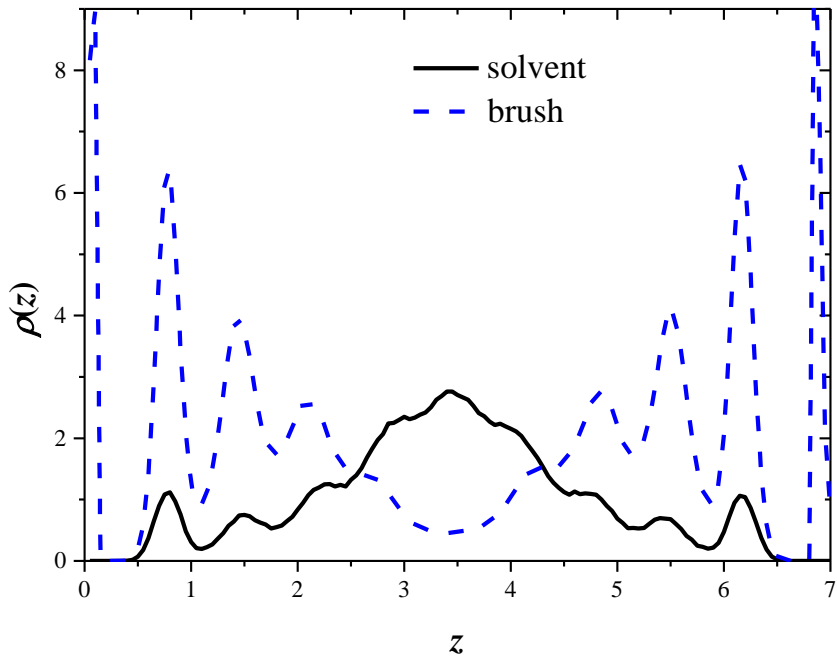

**Figure 2**. Density profiles of the PEG brushes (dashed line, in blue) with polymerization degree $N=7$, and the solvent (full line, in black). The polymer monomers order near the walls, which leads to their structuring, represented here by peaks near the walls, while there is a relatively free (bulk like) fluid made up of solvent molecules at the center of the simulation box. For the case shown in this figure, $L_x^*=L_y^*=7$, $D^*=7$, $v_0^*=1.0$ and $\varGamma^*=1.0$. Note there is very little interpenetration between the brushes. The brush structuring is rather strong because the chains are relatively short.

The profiles of the opposing polymer brushes show very little interpenetration and an almost bulk like concentration of solvent monomers at the center of the simulation box (see Figure 2), which reduces the viscosity of the system. The structuring of the PEG brush monomers, represented by the maxima in Figure 2, coincides with that of the solvent monomers, indicating that even for a relatively large grafting density as that shown in the figure, the solvent is able to reach the membrane. The maxima of both profiles (the solvent's and the brushes') occur at the same positions because the interactions between them are the same (recall $a_{ij}=a_{ii}=78.0$ for all *i* and *j*), as it is befitting for theta – solvent conditions. The strong layering is expected when the chains are relatively short, as in Figure 2, because in such case monomers are more easily arranged that when long chains form the brushes [16-20]. In Figure 3 we present the velocity profile for PEG brushes of polymerization degree $N=14$ and grafting density $\Gamma^*=0.30$ under a shear velocity equal to $v_0^*=0.1$. Clearly, there appears a linear gradient in the center of the channel formed between liposomes, from which the shear rate $\dot{\gamma}$ can be obtained, through the relation $\dot{\gamma} = 2v_0^*/D^*$, as pointed out before. The inset in Figure 3 shows the temperature profile of the complex fluid in the *z*-direction, which indicates the brushes are at the temperature fixed by the thermostat ($k_BT^*=1$), although at large shear rates one expects the brush to "heat up" somewhat [19], as a consequence of the increased dissipation rate.

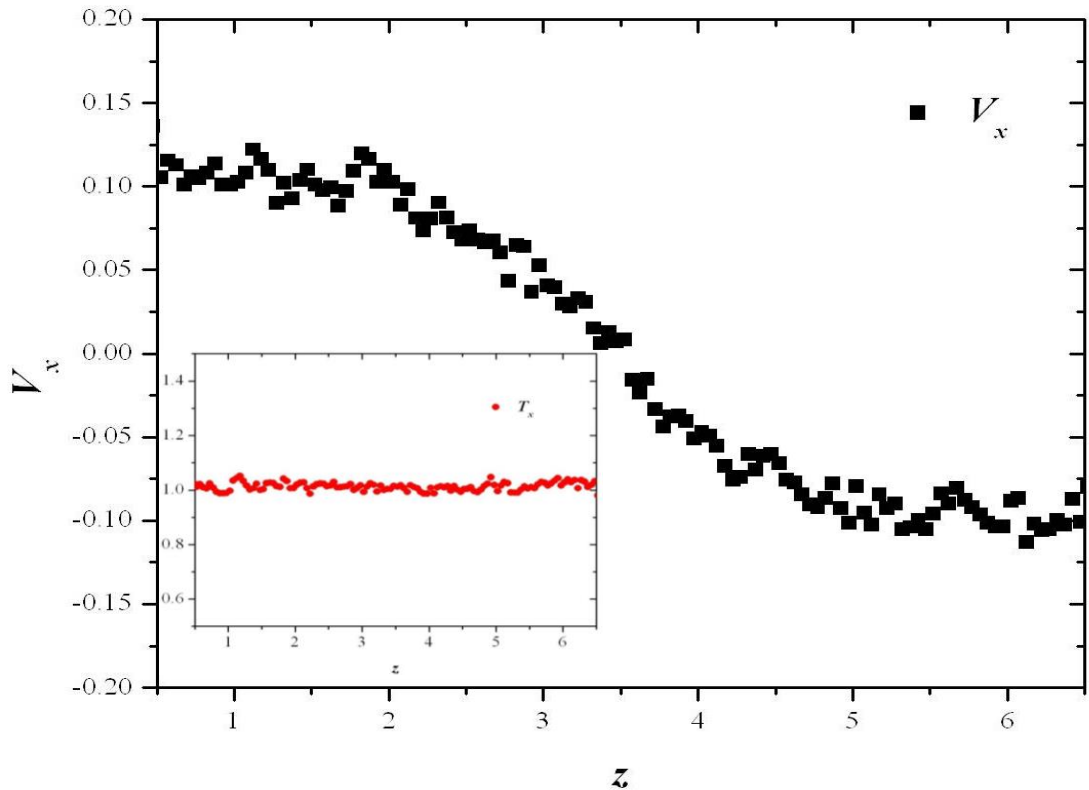

**Figure 3**. Velocity profile at the center of the channel between liposomes for PEG brushes with $N=14$ and $\Gamma^*=0.30$, using $v_0^*=0.1$. The box size is $L_x^*=L_y^*=7$, $D^*=7$. The inset shows the temperature profile ($T_x^*$) in the z-direction. All quantities are reported in reduced units.

Most simulations were performed for a brush grafting density $\Gamma^*=0.30$, except where indicated otherwise, because for this value the average distance between grafted heads on the surface is smaller than their radius of gyration in theta solvent, for all values of $N$ we studied [27]. By doing so one makes sure that the grafted polymer chains are in the brush, rather than the mushroom regime [27], which is important for the situation we are interested in modeling. The polymerization degree was varied in the range $N=1$ up to $N=25$, so that scaling laws could be extracted from the data. Finally, the shear rate was chosen to vary from $\dot{\gamma}=0$ to $\dot{\gamma}=0.30$, except where indicated otherwise, so that we could compare our results with those available in the literature [16-20]. Molecular dynamics simulations like ours solve the equation of motion essentially exactly [33], although the use of soft forces in DPD (see, for example, equation (1)) might limit its applicability to situations where atomistic detail is not important, as is the case for the nanotribology studies that are the focus of this work.

## III. RESULTS AND DISCUSSION

One would like to simulate systems with a large number of particles so that realistic situations can be reproduced, while simultaneously keeping such number manageable to obtain results in a timely fashion. Therefore, we have carried out simulations to quantify the extent of finite size effects in the viscosity of the polymer brushes, if any. Figure 4 shows the mean viscosity of the PEG brushes and the solvent, calculated with equation (6), as a function of the lateral size of the membranes ($L_x^* = L_y^*$), at fixed separation between them ($D^* = 7$), grafting density ($\Gamma^* = 0.30$) and shear rate ($\dot{\gamma} = 0.28$). The results in Figure 4 cover a range that goes from $10^3$ particles up to $10^5$ particles while the relative change in the viscosity amounts to less than 1 %, showing that finite size effects are not important in the calculation of the viscosity, which allows us to make correct predictions using relatively small systems.

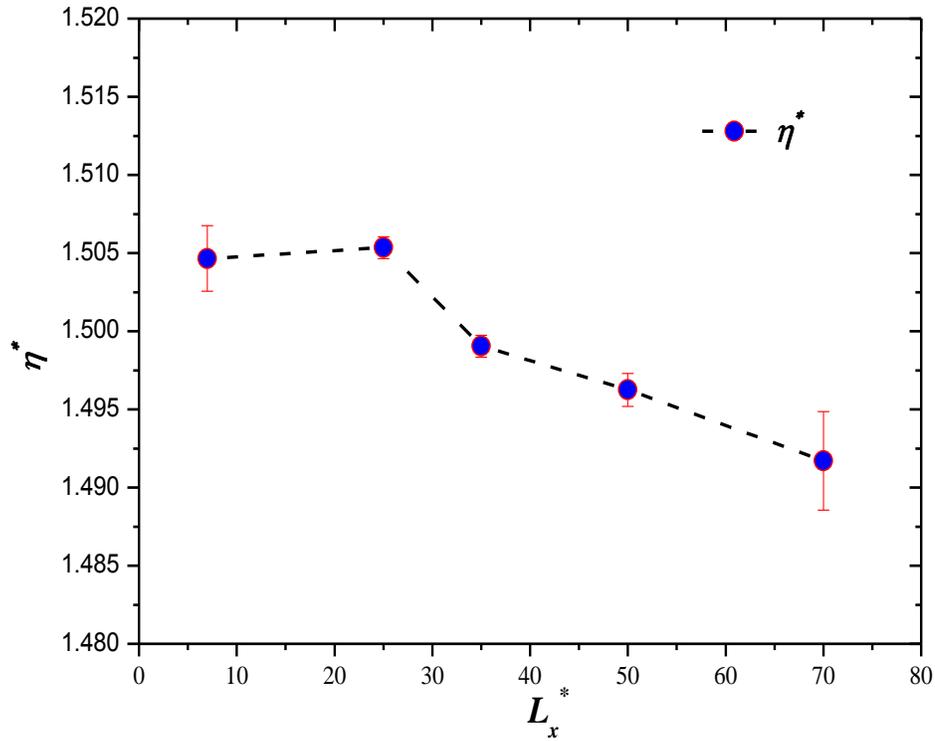

**Figure 4**. Finite size effects in the viscosity, $\eta^*$. The symbol $L_x^*$ represents the size of the simulation box in the $x$ – direction, which is equal to that in the $y$ – direction, $L_y^*$. In all cases, $L_z^* = D^* = 7$, the shear rate is $\dot{\gamma} = 0.28$, and the grafting density is $\Gamma^* = 0.30$. All quantities are reported in reduced units. Lines are only guides for the eye.

Similar results have been found in equilibrium DPD simulations [42] which proved that finite size effects in the interfacial tension between two model fluids were as small as those found in the viscosity and shown in Figure 4, when the appropriate ensemble was used. This behavior is to be attributed to the soft, short − range repulsive nature of the DPD forces (see equations (1)-(3)), although these effects may become non negligible if a long − range interaction is included, such as the electrostatic one. In the remainder of this work we present results obtained using only the smallest simulation box in Figure 4.

Let us now proceed to discuss the results obtained for the mean viscosity of a complex fluid made up of PEG brushes and an aqueous solvent, as a function of the shear rate for different degrees of polymerization (*N*), as seen in Figure 5. The simulations were performed with a cubic box of lateral size $L^*$=7 and for $\Gamma^*$=0.30. For very small values of $\dot\gamma$ (less than $10^{-4}$) the calculation of $\eta^*$ requires of very long simulations due to poor signal − to − noise ratios [19], which makes it very difficult to obtain accurately the zero − shear viscosity, $\eta_0$, from simulations. However, a good estimate of it can be obtained from the extrapolation of the rheology profiles in Figure 5 as $\dot\gamma$ approaches zero. For the shorter PEG brushes (*N*=1 and *N*=3) the fluid displays essentially Newtonian behavior, whereas for the case with *N*=7 the fluid shows shear − thinning. One observes also that as *N* is increased so is the extrapolated $\eta_0$ value, which is a consequence of the combined effects of an increasing brush thickness and a reduced number of solvent particles (to keep the global density $\rho^*$=3). Increasing *N* is equivalent to reducing the separation between the membranes, $D^*$, when the grafting density is kept constant, and yields increasing values of $\eta_0$ [18]. Hence, we have opted to vary *N* rather than $D^*$ in this work. The simulations shown in Figure 5 were carried out for relatively short polymers with the purpose of establishing the minimum polymerization degree at which the fluid started to show shear thinning, which, from Figure 5, is $N=7$. For polymerization degree smaller than 7, it is difficult for the polymer chains to interact with the solvent monomers enough to reduce the viscosity as the shear rate is increased, hence the fluid remains approximately Newtonian ($N=1$ and $N=3$ in Figure 5).

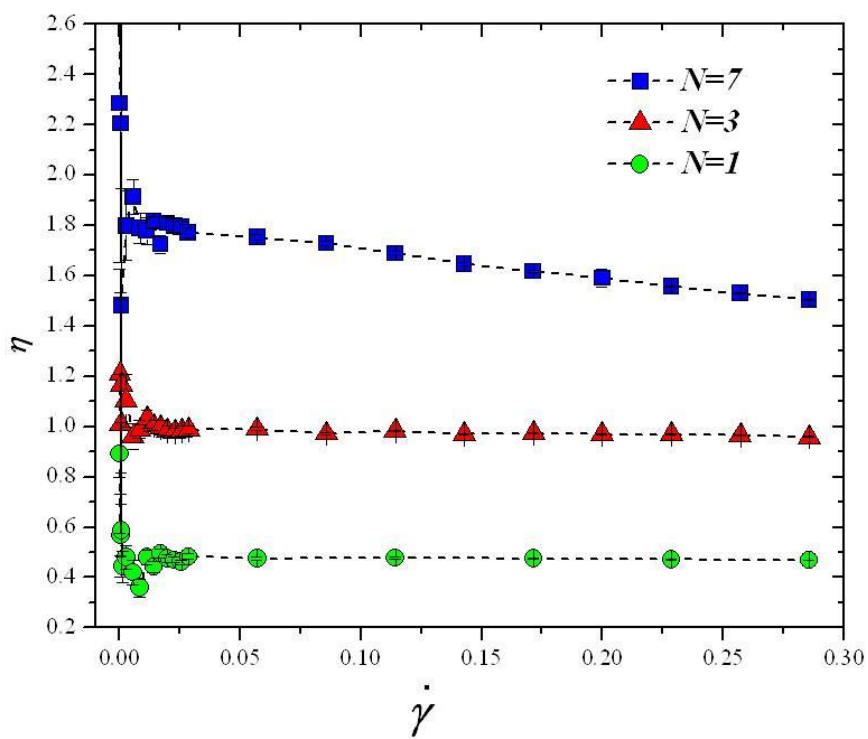

**Figure 5**. Viscosity for a fluid made of polymer brushes of various degrees of polymerization ($N$) and monomeric solvent particles. Notice how the shorter brushes show virtually no shear – thinning, and behave almost like Newtonian fluids. In all cases, $L_x^*=L_y^*=D^*=7$, and the grafting density is $\Gamma^*=0.30$. All quantities are reported in reduced units. Lines are only guides for the eye.

It has been reported that increasing the polymerization degree of PEG brushes on liposomes increases their circulation longevity [28], which would be a desirable aspect to improve drug delivery mechanisms. However, as its degree of polymerization grows, PEG's solubility in aqueous environment is reduced [29], hence there would be a competition between these trends that can be investigated with DPD simulations, varying $N$. Also, it is important to find out if the viscosity behaves qualitatively the same as a function of shear rate, regardless of the value of $N$, so that general conclusions are obtained that can be used to interpret a variety of experiments [24]. Before presenting our results for the effective mean viscosity of systems where PEG brushes have larger polymerization degrees we comment on the behavior of the mean shear stress on the membranes along the direction of the flow as a function of the shear, $\langle F_x(\dot{\gamma}) \rangle$. Following Galuschko and coworkers [43], we find a critical shear rate for our systems, $\dot{\gamma}^*$, as the value of the shear rate when the behavior of $\langle F_x(\dot{\gamma}) \rangle$ changes from linear to sub linear, which depends on $N$, being $\dot{\gamma}^* \sim 8.6 \times 10^{-3}$ for $N$=14, and $\dot{\gamma}^* \sim 2.9 \times 10^{-3}$ for $N$=25. Finding $\dot{\gamma}^*$ is tantamount to finding the Weissenberg number, $We$, since $We = \dot{\gamma}/\dot{\gamma}^*$, and it signal when shear – thinning behavior starts to set in ($We \geq 1$). Figure 6 shows the dependence of the normalized shear stress, $u \equiv \langle F_x(\dot{\gamma}) \rangle / \langle F_x(\dot{\gamma}^*) \rangle$, on $We$ for fluids with brushes with polymerization degrees equal to $N$=14, 20 and 25. The data for the three brushes are seen to collapse reasonably well, especially at low $We$; this is the so called linear regime, $u \sim We$, (dashed blue line in Figure 6). For larger values of the Weissenberg number, we find that the power law $u \sim We^{\kappa}$ is obeyed with $\kappa = 0.69$ (solid black line in Figure 6) for our systems under theta – solvent conditions. Galuschko *et al*. [43] predict $\kappa = 9/13 \approx 0.69$ using scaling arguments for what they call "dry" polymer brushes, which are brushes where the hydrodynamic and excluded volume interactions are screened out, for fairly concentrated solutions and polymer melts, where the corresponding Flory exponent is ν=0.5 [44]. Our simulations predict a $\kappa$ value in excellent agreement with scaling arguments [43] because under theta – solvent conditions and at the concentrations we modeled, the brushes are almost indistinguishable from the solvent, and act as a melt.

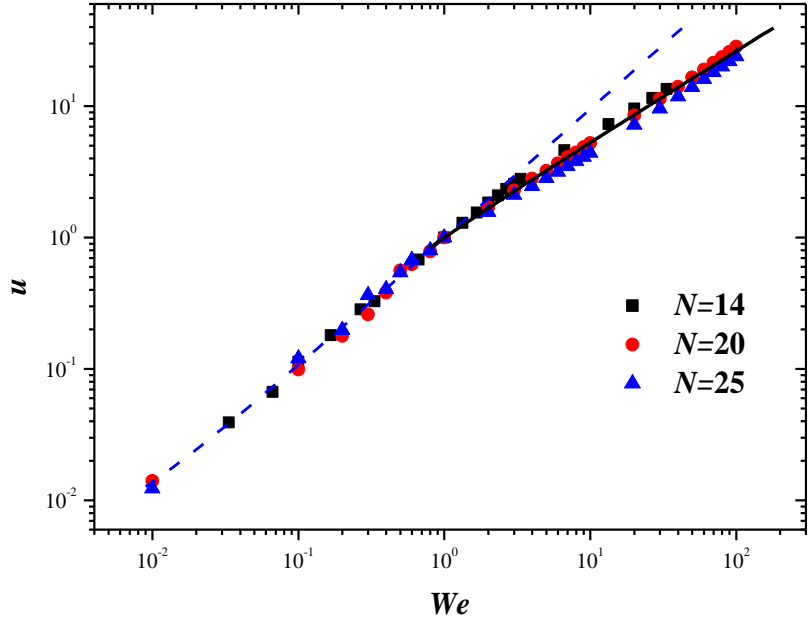

**Figure 6**. Normalized mean shear stress ($u$) on the surfaces in the direction of the flow as a function of the Weissenberg number, $We$, for PEG brushes of three different polymerization degrees, $N$. In all cases, $L_x^*=L_y^*=D^*=7$, and the grafting density is $\Gamma^*=0.30$. The dashed blue line represents the linear regime, $u \sim We$ ($We < 1$), and the solid black line the power law $u \sim We^\kappa$, with $\kappa = 0.69$. See text for details.

Recent molecular dynamics simulations [45] of strongly compressed polyelectrolyte brushes under shear at melt concentrations find that the shear force $\langle F_x(\dot\gamma) \rangle$ that appears in equations (6) and (7) behaves at high shear rate ($\dot\gamma$) as $\langle F_x(\dot\gamma) \rangle \sim \dot\gamma^{0.69}$, in agreement with our predictions in Figure 6, and with the scaling arguments of Galuschko *et al.* [43]. Although there are explicit electrostatic interactions in the work of Spirin and Kreer [45], they argue that their brush and solvent system behaves as a melt of neutral polymers at high *We* because of the immobilization of the counterions. Therefore, the scaling exponent $\kappa$=0.69 we obtained in Figure 6 is very robust.

The effective mean viscosity data of these systems can also be collapsed as a function of *We* if one defines $s = \eta/\eta_0$ [43], for *We* > 1, as seen in Figure 7(a). The value of the zero shear viscosity $\eta_0$ was obtained from extrapolation, using the same procedure as the one described in the discussion of Figure 5. All three cases experience considerable shear thinning as the Weissenberg number is increased, and in this regime the data follow the power law $s \sim We^\zeta$, with $\zeta = -0.31$, see the black line in Figure 7(a). To our knowledge, this is the first time such scaling law has been obtained in theta solvent.

In reference [43] the value $\zeta = -0.43$ was obtained for linear Lennard − Jones brushes in good solvent. A similar value ($\zeta = -0.42$) was found in simulations of bottle − brush polyelectrolytes in good solvent [18]. Hence, polymer architecture appears to be of secondary importance, compared with the quality of the solvent or the polymer concentration. Galuschko and collaborators [43] proposed also the following relation between $\kappa$ and $\zeta$:

$$\kappa - \zeta = 1 \qquad (8)$$

which is clearly fulfilled in our theta solvent simulations. Therefore the relation between the scaling exponents is the same regardless of solvent quality, and is more fundamental than the architecture of the polymer chains or the particular values of the exponents themselves, separately, as is known to be the case for equilibrium scaling exponents [46].

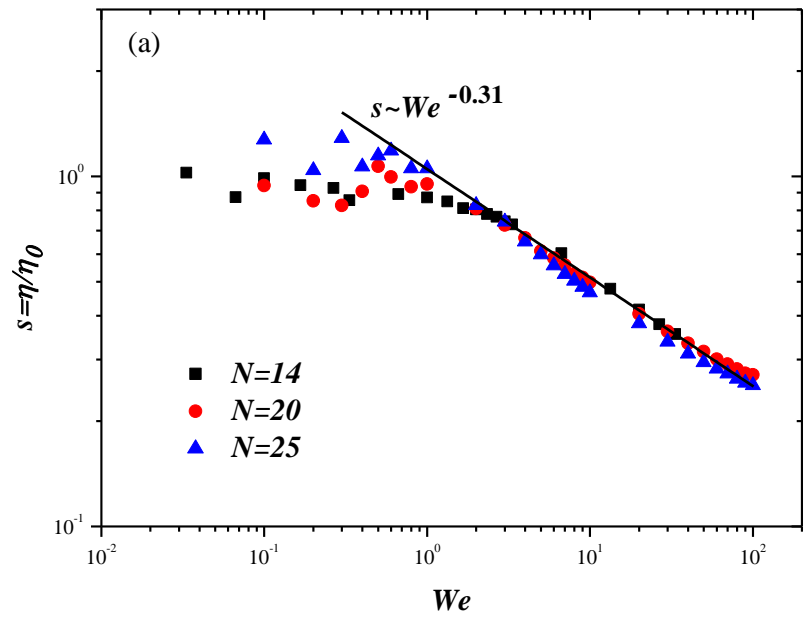

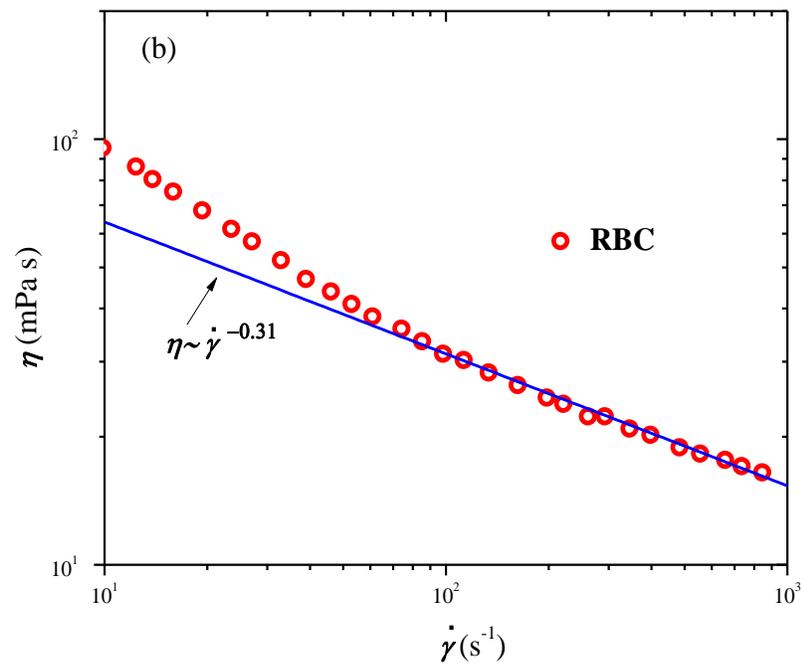

**Figure 7**. (a) Normalized effective mean viscosity $s = \eta/\eta_0$ as a function of the Weissenberg number, *We*, for fluids with PEG brushes of three different polymerization degrees, *N*. In all cases, $L_x^*=L_y^*=D^*=7$, and the grafting density is $\Gamma^*= 0.30$. The line represents the power law, $s \sim We^\zeta$, with $\zeta = -0.31$. (b) Viscosity of a fluid made up of red blood cells (RBC) dispersed in a 40 kDa liposome – dextran medium, as a function of the shear rate [47]. The blue line represents the power law $\eta \sim \dot{\gamma}^{-0.31}$. See text for details.

In Figure 7(b) we have included the high shear rate values of the viscosity of human red blood cells (RBC) in a liposomal suspension [47], which shows evidently shear – thinning. This non – Newtonian behavior is interpreted as being the result of the dissociation of flocculated liposomes at the higher values of the shear rate, an interpretation that is consistent with our predictions in Figure 7(a). The blue line in figure 7(b) represents the power law we have obtained for PEG – grafted surfaces in a theta solvent, which models accurately the results for RBC at high values of the shear rate, providing confirmation for our predictions. Experiments carried out for gels used as vehicles for drug delivering liposomes [48], such as hydroxyethylcellulose (HEC) and a mixture of HEC and an acrylic acid – based polymer (carbopol 974) yield values of the viscosity at high shear rates very close to that found in Figure 7(a), namely $\zeta = 0.30 \pm 0.01$ for HEC and $\zeta = 0.310 \pm 0.009$ for the mixture. The fact that the exponent in these experiments is found to be very close to our prediction suggests that it is of paramount importance to take into account the properties of the solvent (theta conditions), as well as those of the brush (PEG), and lends additional support to these calculations.

Adding polymer brushes to sheared surfaces not only modifies the viscosity of the system but it has been shown to reduce its friction coefficient [10] also. Since this is a parameter that can be directly compared with experiments, we have obtained the friction coefficient of PEG brushes in an explicitly included, theta solvent. Using equation (7) one calculates the mean friction coefficient between the two sliding surfaces. The data in Figure 8(a) show that the friction coefficient $\mu$ at constant PEG grafting density, increases with the shear rate, in agreement with trends found by others [18, 43, 49]; it increases also with $N$ because the overlap between the brushes also grows. Moreover, the $\mu$ values for PEG brushes in theta solvent are somewhat larger than those obtained in simulations of linear brushes in good solvent [17]; this is of course expected because the brush thickness is reduced when it is immersed in a theta solvent, and the contacts between the polymers grow when going from good to theta solvent [2]. Experiments by Klein and collaborators [50] on the friction of polymer brushes in theta solvent compared to good solvent show that the friction coefficient can be larger in the former by up to three orders of magnitude than in the latter. The reason relies on the fact that in theta solvent conditions the brushes can interdigitate and the solvent can penetrate the brushes (see Figure 2), reducing the freely flowing solvent in the center of the channel, thereby increasing the shear force and the friction. Klein and coworkers [50] obtain values for $\mu$ which are close to our predictions, shown in Figure 8(a). When electrostatic interactions are included the values for $\mu$ obtained from computer simulations turn out to be slightly larger and closer to 1 for good solvent conditions at high values of the shear rate [18, 49], due to the added osmotic pressure of the ions. However, in biological systems like those modeled here, the quality of the solvent plays a key role in providing a mechanism that promotes a lower viscosity environment.

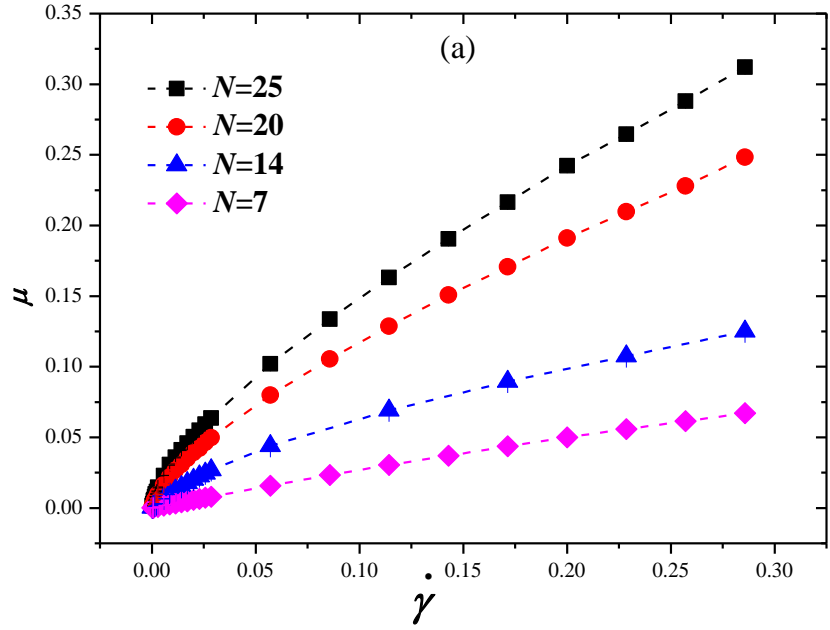

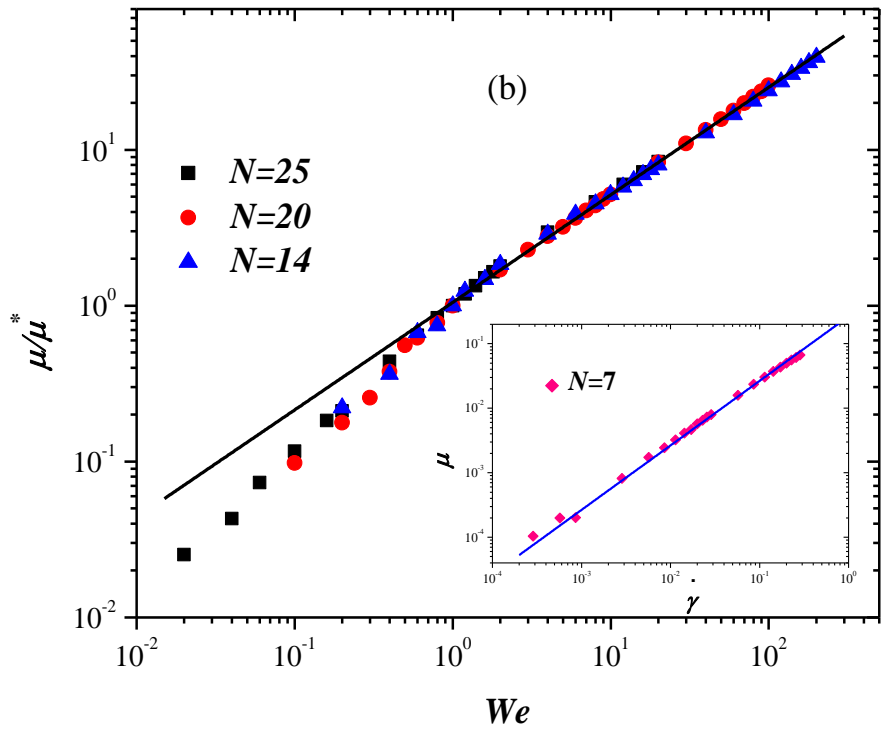

**Figure 8**. (a) Friction coefficient for polymer brushes of various degrees of polymerization ($N$), obtained using equation (7). In all cases, $L_x^*=L_y^*=D^*=7$, and the grafting density is $\Gamma^*= 0.30$. Lines are only guides for the eye. (b) The friction coefficient data shown in (a), normalized by its value at $\dot{\gamma}^*$ for the largest 3 brushes, $\mu/\mu^*$, as a function of the Weissenberg number, $We = \dot{\gamma}/\dot{\gamma}^*$. The black line represents the fit to the power law $\mu/\mu^* \sim We^\kappa$, with $\kappa=0.69$. The inset shows the friction coefficient for the polymer with the smallest degree of polymerization ($N=7$, pink rhombi), with the blue line representing a linear fit, $\mu \sim \dot{\gamma}$.

In Figure 8(b) we show the same data as in Figure 8(a), on logarithmic scales to emphasize power law behavior. For the shortest brush (see inset in Figure 8(b)), made up of chains with polymerization degree equal to $N=7$, there is a linear relation between the friction coefficient and the shear rate, $\mu \sim \dot{\gamma}$, for the entire range of shear rate values used in the simulations (blue line). To see sublinear behavior here one would have to impose larger shear. For the other polymerization degrees shown ($N=14$ to 25) the main panel in Figure 8(b) shows that the data collapse rather well on a single curve when the friction coefficient ($\mu$) is normalized by its value ($\mu^*$) at the start of the shear – thinning regime ($\dot{\gamma}^*$). The black line is the fit to the power law $\mu/\mu^* \sim We^\kappa$, with $\kappa=0.69$. This is the same exponent as the one obtained from the normalized shear stress data, in Figure 6, indicating that the force $\langle F_z(\dot{\gamma}) \rangle$ (see equation (7)) is essentially independent of shear once the brushes have reached a certain length. At the same time, the fact that the exponent is $\kappa=0.69$ at $We > 1$ in all three cases ($N=14$, 20 and 25) shows that $\langle F_x(\dot{\gamma}) \rangle$ (see equation (7)) is responsible for the behavior of the friction coefficient in the high shear rate regime. The same behavior is found for Lennard – Jones brushes [43], namely that the friction coefficient dependence on *We* scales with same exponent as that of the shear stress, although in such case $\kappa=0.57$ because that system is under good – solvent conditions.

## IV. CONCLUSIONS

We have shown that complex fluids with polymer brushes under theta solvent conditions display rheological characteristics that differ in detail from their good solvent counterparts, but that obey nevertheless the same general scaling properties, in particular for the shear stress and the viscosity. The exponent for the shear stress as a function of *We* obtained from our simulations, $\kappa=0.69$, is in excellent agreement with the value predicted using scaling arguments [43] for dense polymer brushes where excluded volume interactions are screened out, as in our model. The viscosity we obtained scales with the shear rate with an exponent equal to $\zeta = -0.31$, which reproduces remarkably well measurements of the viscosity of red blood cells dispersed in a liposome carrying aqueous fluid at high shear rate [47], as well as other experiments [48]. The friction coefficient data as a function of shear rate for different polymerization degrees of the chains making up the brushes were found to collapse on a universal curve whose behavior at $We > 1$ follows the same power law as the shear force, i.e., $\mu \sim We^\kappa$, with $\kappa=0.69$, in agreement with trends found in simulations under good – solvent conditions [43], albeit with a value of $\kappa$ particular to those conditions. Ours are the first simulations, to the best of our knowledge, of the scaling of viscosity and the friction coefficient for systems under theta – solvent conditions.

It is argued that our simulations are useful for understanding the behavior of biopolymer brushes coating drug – carrying liposomes in an aqueous environment that acts as a theta solvent, under shear. Increasing the thickness of the brush through the degree of polymerization or the shear velocity on the surfaces is shown to raise the friction coefficient, which is a deleterious effect for the transport properties of these liposomes. However, when liposomes are covered by PEG bushes in a theta solvent at high shear rates, their flowing characteristics make them optimal carriers for drug delivery, as the polymer brushes imprint them with efficient injectable characteristics (low viscosity at high shear rate) while at the same time providing them with thermodynamically stable ("stealth") mechanisms. These simulations have the additional advantage of including hydrodynamic interactions, as well as the solvent explicitly, and being mesoscopic they reproduce length and time scales comparable with those of environments of biological interest. Therefore, we believe our work should be useful in the improved design of drug – carriers, the rheological characterization of sheared brushes, and in the establishing of general scaling laws for non – equilibrium polymer brushes.

## V. ACKNOWLEDGEMENTS

The authors would like to thank ABACUS, CONACyT grant EDOMEX-2011-C01-165873, for funding. AGG thanks M. A. Balderas Altamirano, C. Carmín, E. de la Cruz, J. P. López Neria and E. Pérez for instructive discussions.